\documentclass[12pt,english]{article}

\usepackage[headheight=15pt]{geometry}
\usepackage{setspace}
\usepackage{mathtools}
\usepackage{enumitem}

\usepackage{sectsty}
\allsectionsfont{\normalsize\raggedright\centering}

\usepackage[]{tocloft}
\addtocontents{toc}{\cftpagenumbersoff{section}}
\addtocontents{toc}{\cftpagenumbersoff{subsection}}

\makeatletter
\renewcommand{\@cftmaketoctitle}{}
\makeatother

\usepackage[]{footmisc}
\usepackage[authoryear, round]{natbib}
\usepackage{url}

\makeatletter
\def\url@leostyle{%
    \def\UrlFont{\sf}}{\def\UrlFont{\small\ttfamily}}
\makeatother
\bibpunct
{(} %the opening bracket symbol
{)} %the closing bracket symbol
{,} %the punctuation between multiple citations
{a} %the letter `n' for numerical style, or `s' for numerical superscript style, any other letter for author-year
{} %the punctuation that comes between the author names and the year
{;} %the punctuation that comes between years or numbers when common author lists are suppressed

\usepackage{authblk}

\usepackage{graphicx}
\usepackage{babel}

\usepackage{amsmath}
\usepackage{amsfonts}
\usepackage{amssymb}

\usepackage{framed}

\usepackage{xifthen}

\usepackage{fancyhdr}

\begin{document}

\title{Transfer of quantum information in teleportation}
\author{Lev Vaidman}
\affil{{\small  School of Physics and Astronomy, Tel-Aviv University, Tel-Aviv 69978, Israel}}
\date{}

\maketitle

%% no headers on the first page:
\thispagestyle{empty}

\begin{abstract}
The controversial issue of information transfer in the quantum teleportation procedure is analyzed in the framework of the many-worlds interpretation of quantum mechanics. It is argued that quantum information, considered as a measurable property for an observer in a particular world, is transferred in a nonlocal way in the teleportation process. This, however, does not lead to an action at a distance on the level of the universe which includes all parallel worlds. The alternative approach of Deutsch and Hayden is discussed. 
\end{abstract}
%%%%%%%%%%%%%%%%%%%%%%%%%%%

%%%%%%%%%% Insert the texts which can accomdate on firstpage in the tag "fmtext" %%%%%
\maketitle

\section{Introduction}

Recently, we have witnessed a rapid development of quantum information science fueled by a quantum technology revolution which made experimental implementation of many theoretical ideas possible. Philosophical analyses of quantum concepts, which were introduced at the birth of quantum theory but never reached a consensus, become more relevant than ever. Here, I analyze arguably the most bizarre quantum information protocol: quantum teleportation, a transfer of a quantum state with surprisingly small resources.

When Asher Peres, coauthor of the teleportation paper (Bennett et al. 1993) was asked by a reporter if quantum teleportation could teleport the soul as well as the body, he answered: ``No, not the body, just the soul.'' What is transferred in the teleportation protocol, and how, is still the matter of controversy. The indistinguishability of the quantum particle led Saunders (2006) to ask the question: ``Are quantum particles objects?''. But this indistinguishability is what made teleportation possible: the particle (the ``body'') is not moved in the teleportation protocol. It is the quantum state of a particle (the ``soul'') in one site that is transferred to a particle in another site.  The genuine indistinguishability of elementary particles makes it possible to create exactly the same particle in a remote location. 

People are not teleported today from one city to another, and it is safe to say that it will never happen, but the teleportation protocol has become one of the cornerstones of quantum information. The mathematics of teleportation is uncontroversial, but we still need to gain understanding of the paradoxical features of this process (see Vaidman 1994a): how one can send a quantum state, specification of which requires a large amount of  information, by  sending only a tiny amount of information through a classical channel: two bits instead of two angles for sending a qubit (Bennett et al. 1993), or two real numbers
instead of two real-valued functions for sending a quantum state of a 
continuous variable (Vaidman 1994b). 

The approach of Deutsch and Hayden (2000) to the question of information flow in the teleportation protocol created a large controversy regarding the concept of ``quantum information'', see Duwell (2001, 2003), Timpson (2005), Wallace and Timpson (2007), Deutsch (2011), Lombardi et al. (2016), Lopez and Lombardi (2018), Bedard (2023).
Timpson (2006, 2013) argued that the way out of this conundrum is to realize that ``information'' is an abstract concept, and it is a mistake to take the view that ``something travels from Alice to Bob in teleportation...  in a spatio-temporally continuous fashion''. I, however, argue that it is possible to have a coherent spatio-temporal picture of quantum information transfer in teleportation. The core of the controversy is that relevant concepts are often understood in different ways. Although I, as Deutsch and Hayden, perform the analysis in the framework of the many-worlds interpretation (MWI) (Everett 1957, Vaidman 2021), my conclusions are very different: the nonlocality of Everett's worlds is the basis of the teleportation of quantum information.

Another line of research that attempts to explore the meaning of quantum teleportation might be mentioned. It culminated in a paper ``Classical Teleportation of a Quantum Bit'' (Cerf et al. 2000). 
 A natural interpretation of such a title is the transfer of a quantum state from Alice to Bob, both of whom have quantum capabilities but do not have a quantum channel.
The authors write instead:
``Classical teleportation is defined as a scenario where the sender is given the classical description of an arbitrary quantum state while the receiver simulates any measurement on it.'' It is surprising that this task can be achieved with shared randomness instead of an entanglement channel and a very small amount of transferred classical information. However, this scenario does not provide what is promised in the title of the paper: a quantum state was not teleported in this procedure.
In the teleportation scenario, the sender receives a qubit, not its classical description.  More importantly, the output should be a qubit. An agent gives a qubit to Alice and then goes to Bob at a remote location to test the qubit with his measuring device. For a proper teleportation, the agent's verification measurement of the state of the qubit should have probability 1 to succeed. This cannot be done without entanglement. 

 I also would not consider the proposal of Braunstein (1966) as a qubit teleportation because it is fully quantum. The ``classical bits'' sent between Alice and Bob are, in fact, entangled qubits, so it is not surprising that we transfer a qubit by sending two qubits. 
 
 The experimental implementation of teleportation of a qubit was first attempted by Bouwmeester et al. (1997) and achieved by Pan et al. (2003). The experiment by Boschi et al.  (1998), 
performed at the same time, had as an input the classical information of the qubit and not a quantum state, so I do not consider it as an implementation of the qubit teleportation as it defined in the original proposal by Bennett et al. (1993).

The quantum teleportation I analyze here is the process in which Alice obtains an unknown qubit, sends only two bits of classical information to Bob, and the qubit is reconstructed at Bob’s remote site on some other quantum system. To avoid ambiguity, I adopt the following definition:
\begin{quote}
{\bf Location of Quantum Information:} The location of the information about a qubit is the location of the physical systems that allow for the creation of the qubit.
\end{quote}
In this paper, I analyze the location of information about the qubit during the teleportation process.
At the beginning, the information is solely at Alice’s site; at the end of the procedure, the information is solely (due to the no-cloning theorem) at Bob’s site.
 
Here is the plan for the rest of the paper:
In  Section 2, I analyze the location of information in the qubit encoding procedure to clarify the concept of  ``information about the qubit''.  Section 3 analyzes teleportation of a qubit. An alternative approach by Deutsch and Hayden is discussed in Section 4. Section  5 summarizes the results of the paper.

\section{Copying quantum information}

Before discussing information transfer in teleportation, I start with the analysis of the ``location of information'' in a simpler (unitary) process of copying a qubit to a composite system consisting of several two-state systems, i.e., systems that have (at least) two orthogonal states.
 We cannot clone a qubit, however, as we have learned from Shor's method of error correction (Shor 1995), we can create some redundancy by performing a unitary encoding of the qubit in  several two-state systems. 
 Here an encoding of a qubit
in a state of four particles in an error prevention code (Vaidman et al. 1996)  is considered.

The encoding and decoding procedure is described in Fig.~1.
 Although all procedures described in the paper are fine also for a general case of an entangled qubit, for simplicity, we will assume that we encode (and  later teleport) a qubit in a pure state \begin{align}\label{qubit}
|\psi\rangle_{{\rm qubit}}=\left({\sqrt p}~|0\rangle+ e^{i\varphi} \sqrt{1-p} ~ |1\rangle\right), ~~~~~  p \in [0,1],~~\varphi \in [0,2\pi).
\end{align}

\begin{figure}
  \includegraphics[width=\linewidth]{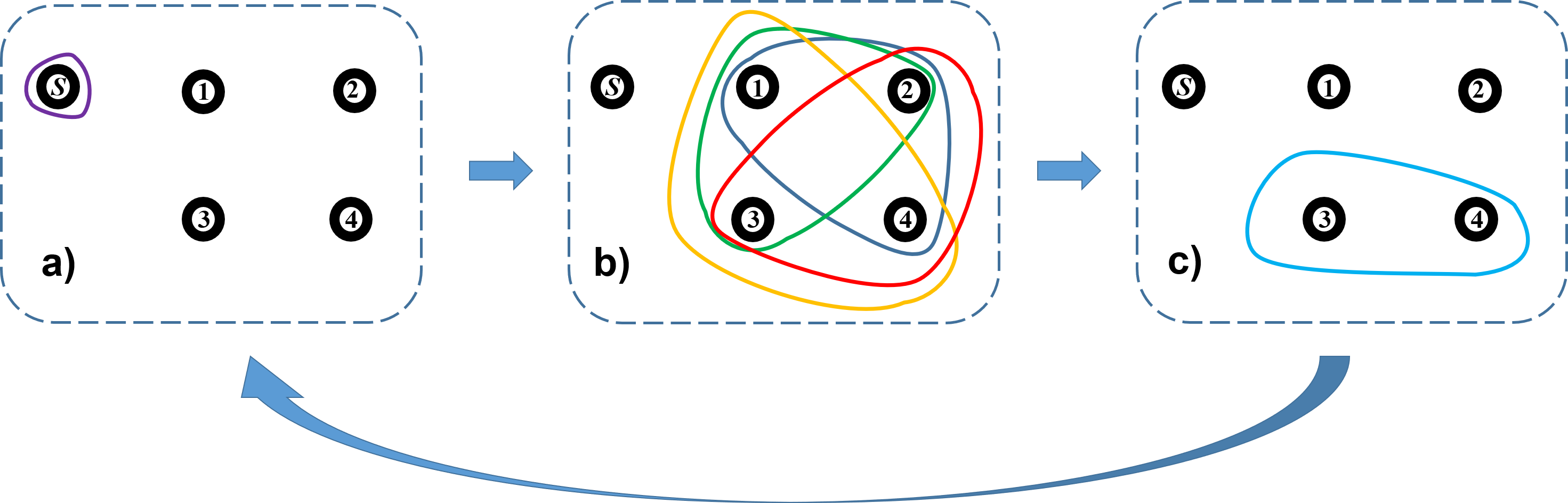}
  \caption{{\bf Information transfer of the qubit in encoding and decoding procedure.} Coloured closed curves show where the full information about the qubit is present. a) At the beginning the qubit (\ref{qubit}) is fully and solely in system $S$. b) After encoding by transformation (\ref{transf}), the information about the qubit is present in any set of three out of the four particles described by state (\ref{4qubitenc}). c) After the interaction between the pair of particles 3 and 4 and particle 2, the information about the qubit is in particles 3 and 4 described by state (\ref{2qubitenc}). The final swap moves the qubit back to $S$.}
  \label{fig:setup}
\end{figure}

We start by preparing two pairs of maximally entangled particles 1,2 and 3,4. Then we apply the following unitary transformation (swap) between our qubit
 and the four particles:
\begin{align}\label{transf}\nonumber
 \frac{1}{2}|0\rangle(|0\rangle_1|0\rangle_2+|1\rangle_1|1\rangle_2)(|0\rangle_3|0\rangle_4+|1\rangle_3|1\rangle_4)\rightarrow \frac{1}{2}|0\rangle(|0\rangle_1|0\rangle_2+|1\rangle_1|1\rangle_2)(|0\rangle_3|0\rangle_4+|1\rangle_3|1\rangle_4),\\
\frac{1}{2}|1\rangle(|0\rangle_1|0\rangle_2+|1\rangle_1|1\rangle_2)(|0\rangle_3|0\rangle_4+|1\rangle_3|1\rangle_4) \rightarrow \frac{1}{2}|0\rangle(|0\rangle_1|0\rangle_2-|1\rangle_1|1\rangle_2)(|0\rangle_3|0\rangle_4-|1\rangle_3|1\rangle_4).
\end{align}
As a result, our original system will not have information about the qubit (\ref{qubit}), but the four other particles will be in a state
\begin{align}\label{4qubitenc}
 \sqrt p \frac{(|0\rangle_1|0\rangle_2+|1\rangle_1|1\rangle_2)(|0\rangle_3|0\rangle_4+|1\rangle_3|1\rangle_4)}{2} +e^{i\varphi}
 \sqrt{1-p}\frac{(|0\rangle_1|0\rangle_2-|1\rangle_1|1\rangle_2)(|0\rangle_3|0\rangle_4-|1\rangle_3|1\rangle_4)}{2},
\end{align}
which encodes the information about the qubit. 
If we consider the four particles as a single composite system, then (\ref{4qubitenc}) represents the original qubit.

Any three of the four particles are not described by a state corresponding to a qubit, but, somewhat surprisingly, they allow for a full reconstruction of the qubit, so, according to my definition, they carry information about the qubit.
 Indeed, assume that, say, particle 1 is lost. Conditioned on the state 
 $\frac{1}{\sqrt{2}}(|0\rangle_3|0\rangle_4-|1\rangle_3|1\rangle_4) $
 of the pair of particles 3 and 4, we can change the relative phase of the entangled particles 1 and 2 from $\frac{1}{\sqrt 2}(|0\rangle_1|0\rangle_2-|1\rangle_1|1\rangle_2)$ to $\frac{1}{\sqrt 2}(|0\rangle_1|0\rangle_2+|1\rangle_1|1\rangle_2)$ by interaction only with particle 2. This will lead to ``decoupling'' of the pair of particles 1 and 2, they will be in a product state with the pair of particles 3 and 4 that will now encode the qubit:
 \begin{align}\label{2qubitenc}
\sqrt p ~\frac{|0\rangle_3|0\rangle_4+|1\rangle_3|1\rangle_4}{\sqrt 2} +e^{i\varphi} \sqrt {1-p} ~\frac{|0\rangle_3|0\rangle_4-|1\rangle_3|1\rangle_4}{\sqrt 2}
.
\end{align}
 Another unitary swap can put the qubit back on our system.
   Note that although any three of the four particles in state (\ref{4qubitenc}) contain full information about the qubit, we cannot obtain any information about the qubit from any single particle. (We can get some information about $p$ from pair 1,2 or from pair 3,4.)
  
The procedure for encoding information described in this section does not involve any kind of action at a distance: local coupling between the particles results in the transfer of information between them. There is some kind of nonlocality feature in the fact that when particles 1-4 are moved to spatially separated sites, the information is encoded in a nonseparable way: we cannot get full information in a single local site without quantum channels between the sites.

The entanglement of the initial state of four particles allows another nonlocality feature. The alternative to the swap operation (\ref{transf}) is the time reversal of the procedure to obtain the qubit from interactions with particles 2,3, and 4. At the end of the procedure, particle 1 contains some information about the qubit since particles 1, 2, and 3 have full information about the qubit, and particles 2 and 3 do not have full information about the qubit. This is a nonlocal feature because particle 1 had no interaction with any quantum system after the time we obtained the qubit.

 \section{Teleportation of a qubit}
 
 In the teleportation of a qubit (Bennett et. al. 1993), spatially separated Alice and Bob share a maximally entangled pair of particles 1 and 2:
 \begin{align}\label{EPR}
  |\psi\rangle_{\rm EPR}=\frac{1}{\sqrt 2 }(|0\rangle_1|0\rangle_{2} +|1\rangle_1|1\rangle_{2}).
\end{align}
Alice receives a qubit (\ref{qubit}) on a system $S$ and performs a Bell measurement on her qubit $S$ and particle 1, see Fig. 2. 
The Bell measurement can be made by performing two consecutive measurements. The first is the measurement of the modular sum of variables $s$ of the system and particle 1 with eigenstates $|0\rangle$ and $|1\rangle~$  ($|0\rangle_1$ and $|1\rangle_1$), and eigenvalues defined by $s|i\rangle =i|i\rangle~~$ ($s_1|i\rangle_1 =i|i\rangle_1$). A conceptually simple method for such a measurement is consecutive couplings of another qubit with $s$ and $s_1$. The auxiliary qubit starts with 0 and is flipped on condition $s=1~$, and then conditionally flipped again if $s_1=1$. The result of the measurement is obtained when the auxiliary qubit is observed by a macroscopic device and is written in bit 3:

 \begin{align}\label{BellM1}
{ (s+s_1)\hspace{-8pt}\mod2 } \equiv b_3.
\end{align}
The second measurement is of the same kind, but in a different basis. The result will be written in bit 4:
 \begin{align}\label{BellM2}
{ (\tilde s+\tilde s_1)\hspace{-8pt}\mod2} \equiv b_4,
\end{align}where 
  \begin{align}\label{tilde}
 |\tilde 0\rangle \equiv\frac{1}{\sqrt 2 }(|0\rangle +|1\rangle),~~~~~~~
 |\tilde 1\rangle \equiv\frac{1}{\sqrt 2 }(|0\rangle -|1\rangle), ~~~~~~ \tilde s|\tilde i\rangle =i|\tilde i\rangle,
\end{align}
and similar definitions for particle 1.  (The mathematics is the same as in measurements of nonlocal variables, see Aharonov et al. (1986) ).

\begin{figure}
 \centerline {\includegraphics[width=10 cm]{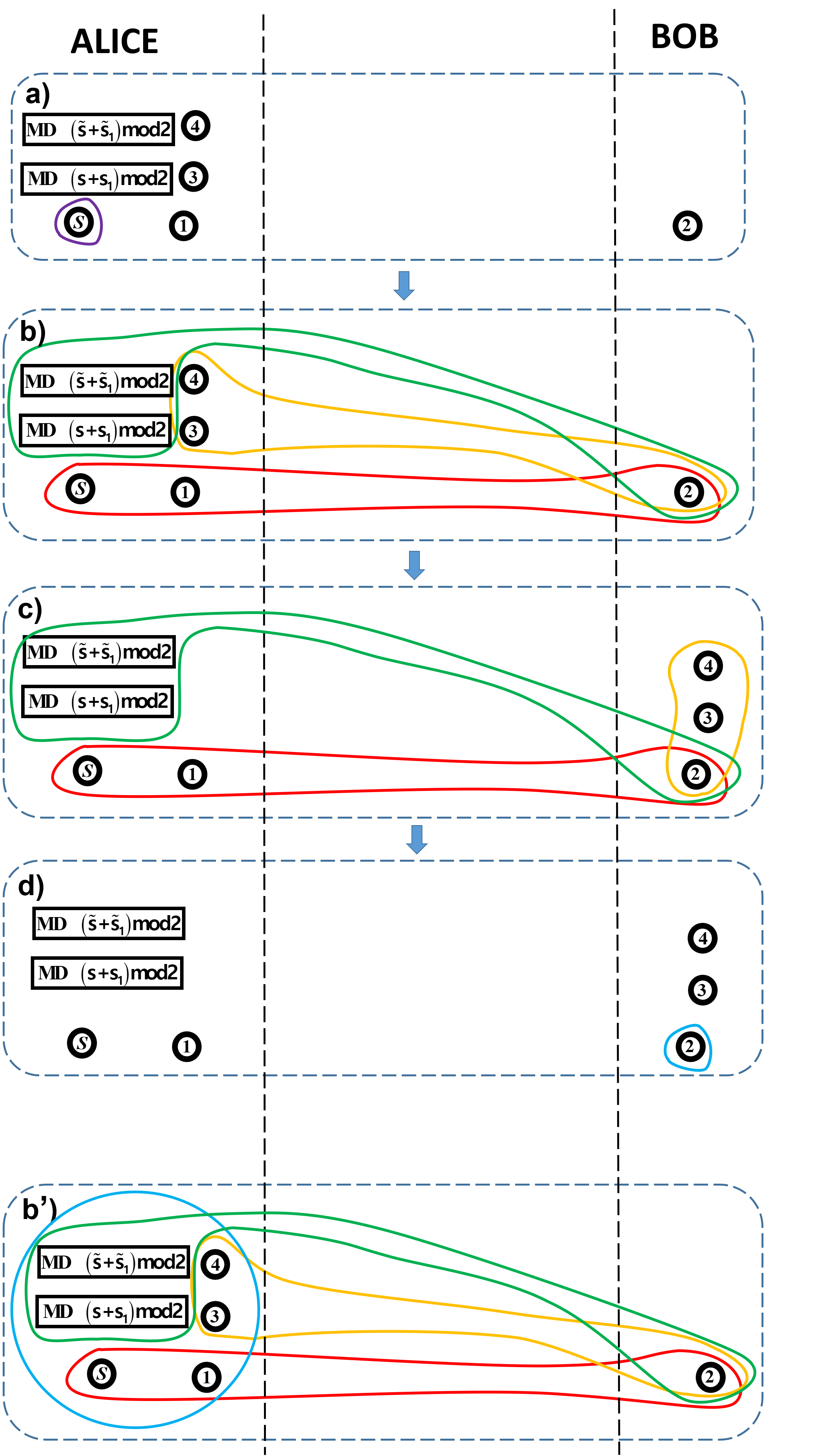}}
  \caption{{\bf Information transfer in the teleportation of a qubit.} Colored closed curves show where complete information about the qubit is present.  Rectangular boxes represent macroscopic measuring devices. a) At the beginning, the qubit (\ref{qubit}) is fully and exclusively located in the system $S$. b) Location of information about the qubit after performing the Bell measurement on the system and particle 1 within every world corresponding to all possible outcomes (\ref{telep1}-\ref{telep4}). In all worlds we need particle 2 to reconstruct the qubit. (Not all sets are shown, e.g., particle 2, system 3, and the measuring device of $\tilde s_4$.) c) The carriers of the results of the Bell measurement, particles 3 and 4, are moved to Bob. d)~After unitary transformation of the state of particle 2, conditioned on the information brought by systems 3 and 4, particle 2 at Bob's site, and only particle 2, carries the qubit.  b')~If we consider all worlds together (and not separately, as was done in (b)), then the complete information about the qubit is also present in all systems coupled to the system $S$ or to the systems coupled to the systems that coupled to $S$, etc. within the light cone of the Bell measurement event (blue circle).}
  \label{fig:setup1}
\end{figure}

The world will be split into four worlds according to the outcomes of the macroscopic measuring devices, $b_3$ and $b_4$. In these worlds, the quantum states of all microscopic systems involved, including the carriers of the results of the measurements to be sent to Bob,  are: 
  \begin{align}\label{telep1}
|0\rangle_3|\tilde 0\rangle_4 ~\frac{1}{\sqrt 2 }(|0\rangle|0\rangle_1 +|1\rangle|1\rangle_1) ({\sqrt p}~|0\rangle_2+ e^{i\varphi} \sqrt{1-p}~ |1\rangle_2),
\end{align}
 \begin{align}\label{telep2}
|0\rangle_3|\tilde 1\rangle_4 ~\frac{1}{\sqrt 2 }(|0\rangle|0\rangle_1 -|1\rangle|1\rangle_1) ({\sqrt p}~|0\rangle_2- e^{i\varphi} \sqrt{1-p} ~|1\rangle_2),
\end{align} 
 \begin{align}\label{telep3}
|1\rangle_3|\tilde 0\rangle_4~ \frac{1}{\sqrt 2 }(|0\rangle|1\rangle_1 +|1\rangle|0\rangle_1) ({\sqrt p}~|1\rangle_2+ e^{i\varphi} \sqrt{1-p} ~|0\rangle_2),
\end{align}
\begin{align}\label{telep4}
|1\rangle_3|\tilde 1\rangle_4 ~\frac{1}{\sqrt 2 }(|0\rangle|1\rangle_1 -|1\rangle|0\rangle_1) ({\sqrt p}~|1\rangle_2- e^{i\varphi} \sqrt{1-p} ~|0\rangle_2).
\end{align}

We can see explicitly that at this stage, for every result of these measurements (which can be seen in the states of the information carriers $|i\rangle_3 $ and $|\tilde i\rangle_4 $), the information about the qubit is encoded in particle 2. The quantum states of all other systems are independent of the parameters $p$ and $\varphi$ that describe the qubit.

In each world, particle 2 together with the identity of the world is enough to reconstruct the qubit. The identity of the world can be learned from bits 3 and 4. In world $(0_3,\tilde 0_4)$, particle 2 as is, represents the qubit, see (\ref{telep1}). In world $(0_3, \tilde 1_4)$, we should introduce the phase $\pi$ to state $|1\rangle_2$,  see (\ref{telep2}). In world $(1_3, \tilde 0_4)$, we should flip the qubit $|0\rangle_2 \rightarrow |1\rangle_2, ~ |1\rangle_2 \rightarrow |0\rangle_2$, see (\ref{telep3}).  Finally, in world $(1_3, \tilde 1_4)$, we should introduce phase $\pi$ to state $|1\rangle_2$ and flip the qubit, see (\ref{telep4}).  

The identity of the world can be read from the (macroscopic) measuring devices, or from systems 3 and 4 which are sent to Bob, or from the Bell states of the composite system which includes the system $S$ and particle 1. Note that all systems which carry information about the identity of the world are located, at this stage, at Alice's site, see Fig. 2b. Without this information, Bob cannot learn anything about the qubit. Of course, it cannot be otherwise, because a superluminal signalling is impossible. Only after the carriers of the information about the world identity, systems 3 and 4, are moved to Bob, Fig. 2c, he reconstructs the qubit on particle 2, Fig. 2d.

 After the Bell measurement,
  within every world corresponding to a particular outcome of the measurement, the information about the qubit is present in sets of systems all of which include particle 2 located far away from system $S$, the original carrier of the qubit, see Fig. 2b. So, although superluminal signaling is not present here, we do have some superluminal feature in the teleportation procedure. 
  It is a counterpart of the superluminal property of the collapse of the wavefunction (spooky action at a distance). 
   Just before the Bell measurement, the information about the qubit is present solely at Alice's site.
  Immediately after, in every world created by the Bell measurement, full information about the qubit is not present at Alice's site, but spread out among spatially separated particles. (This is also true just after completion of the measurement (\ref{BellM1}).)   At every world, the information about the qubit is distributed across several systems including one system (particle 2) in a spacelike (relative to the Bell measurement event) location. 
 Particle 2 has to be supplemented with some information to reconstruct the qubit. This additional information is brought by particles 3 and 4, see Fig. 2c-d.
 
 My claim that the Bell measurement creates some kind of superluminal action needs clarification. The world-splitting view prevents the concept of diachronic identity between a world now and a world at a later time. 
  The world in which the Bell measurement was performed evolves into a multitude of worlds.
Therefore, it is not obvious how to discuss the time evolution within a world when splitting occurs. One attempt to do this, is to look at the four-dimensional ``worm view'' of agents (Wilhelm 2023). However, the MWI does not have worms as part of the ontology. The experience of an agent in a particular time supervenes on its quantum state and he cannot ask what outcome he will see in the future because he will evolve into multiple selves, each one seeing a different result. So, in order to keep the worm view, we need to add an ontology (such as the ``minds'' of Albert and Loewer (1988)) which would make the MWI much less attractive.

I propose to consider the time evolution within the MWI by running the history of a particular world forward in time. This history is recorded in the world's quantum state at a particular time. In the quantum state of a world with a particular outcome of the Bell measurement after the teleportation procedure with a particular records we can see a ``four-dimensional'' agent with a well-defined evolution backward in time from Bob's manipulation of particle 2, to backward motion of the carriers of the result of the Bell measurement to a particular outcome of Bell measurement, etc.
We can analyze this history forward in time. It becomes the usual time evolution that we experience: preparation of Bell measurement, obtaining one of the results, etc. It is not a unitary evolution. It can be described by the usual Von Neumann story with two processes, the evolution according Schr\"odinger equation and collapses.  We can ask questions about the presence of superluminal phenomena in this evolution. The nonlocality feature of the teleportation procedure is the following temporal sequence: presence of a qubit in one location (Alice), immediate disappearance of the information about the qubit from this location and its appearance in a state of several spatially separated particles. Contrary to the action at a distance of the collapse process, this superluminal feature is not an actual process in space-time, but a property of records at a particular time that describe the history of a particular world.  

 In the teleportation procedure, the nonlocality feature is stronger than in the encoding procedure. In addition to the nonseparability feature of the information distributed in two locations at some stages, we have an action at a distance when we consider histories within separate worlds: a measurement on Alice's site transfers (superluminally) some part of the information to particle 2 at Bob's site. We do not get any amount of locally accessible information in particle 2 in a superluminal way, but we know that the part of the information is present there because, in every world, the qubit cannot be recovered without particle 2.

 Let me summarize the presented view of the superluminal phenomena within a world with a particular outcome of the Bell measurement. The world with separate in space entangled particles is a nonlocal (nonseparable) entity. The local Bell measurement splits the world into worlds that have different properties in remote particle 2 due to the initial entanglement between particles 1 and 2. This is an effective action at a distance within each world. However, the mixture of four states of particle 2, corresponding to the four worlds with different outcomes of the Bell measurement, is identical to the original mixed state of particle 2 in the Einstein-Podolsky-Rosen  state.  Thus,  there is no superluminal effect on the physical level of all worlds together.

 \section{Deutsch-Hayden approach  }

The Deutsch-Hayden (2000) analysis of information flow was also performed in the framework of the MWI and considered all worlds together, so it is not surprising that in the abstract they made a similar claim:

\begin{quote}
 Measuring or otherwise interacting with a quantum system $S$ has no effect on distant systems from which $S$ is dynamically isolated, even if they are entangled with $S$.
 \end{quote}
However, the other message of their work is very different from my approach:
 \begin{quote}
All information in quantum systems is, notwithstanding Bell’s theorem, localized. ... Using the Heisenberg picture to analyse quantum information processing
makes this locality explicit, and reveals that under some circumstances (in particular,
in Einstein-Podolsky-Rosen experiments and in quantum teleportation), quantum
information is transmitted through `classical’ (i.e. decoherent) information channels.
  \end{quote}
So, Deutsch and Hayden claim that the full information about the qubit is stored in particles 3 and 4. I, however,  have shown that in every world (corresponding to a particular outcome of a Bell measurement) we cannot reconstruct the qubit without particle 2. Even if we consider all worlds together, particles 3 and 4 alone are not enough to reconstruct the qubit. 

How can we reconcile the differences in the analyses? The following quote from Deutsch and Hayden makes the differences clear.
\begin{quote}
When analyzing information flow in the
Schrodinger picture, it is essential to realize that it is impossible
to characterize quantum information at a given instant using the state vector alone.
\end{quote}
In contrast, my starting point is that {\it everything} is described by the quantum state. It is a complete description at a particular time. Experiences of all conscious beings in all worlds at that time supervene on the wavefunction of the universe at that time. Deutsch and Hayden apparently want to describe more. They write: 
\begin{quote}The latter [Schr\"odinger picture] is optimized for predicting the outcomes
of processes given how they were prepared, but (notoriously) not for explaining how
the outcomes come about...
\end{quote}
They expect the quantum picture to explain the situation not just now, at time $t$ but also at other times. Indeed, the roots of their approach stem from Gottesman (1998) analysis, which dealt with practical aspects of quantum computation. A particular intermediate state of the computer makes sense as a part of the computation when a computer program is given. We need some information about the history of the system to give the meaning of the state as a computational step. The Heisenberg picture includes this history. This can be seen from the continuation of Deutsch and Hayden's writing.
\begin{quote}
To investigate where the information is located, one must also take into account how that state came about. In the Heisenberg picture, this is taken care of automatically, precisely because the Heisenberg picture gives a description that is both complete and local.\end{quote}
 
 This ``complete and local'' description is an example of a local realistic model of quantum mechanics which must exist,  as argued by Brassard and Raymond-Robichaud (2019), since quantum mechanics is a nonsignalling theory. The description based on ``descriptors'' introduced by Deutsch and Hayden,  extensively analyzed by Timpson (2013) and further developed by several authors (Hewitt-Horsman and Vedral 2007; Waegell 2018; Raymond-Robichaud 2021), comes for a high complexity price (see Bedard 2021). 
  The description is local because it is based on local descriptors affected by local interactions. By adding the assumption of a known initial state, we obtain a picture in which all information is about local facts (interactions), and it is stored locally in local descriptors. However, the interaction of a particle with every quantum system increases the Hilbert space of the descriptors (Bedard 2021), so the description becomes very complex. Conditioning on a particular outcome when the interaction is a measurement with a macroscopic measuring device, i.e., considering descriptors in a particular world, (see Kuypers and Deutsch 2021) reduces the complexity, but only a little. 
 
 In the language of Raymond-Robichaud (2021), the local description is based on ``noumenal'' states. Given an ensemble of universes in a particular quantum Schr\"odinger state at a particular moment and external omnipotent quantum devices, we can perform tomography and find the quantum state of the universe. In contrast, the access to the ensemble of universes at a particular time is not enough to specify the description with descriptors, we need the access during the history of the universe. Moreover, even then, we apparently will not be able to find all the descriptors. This is because the descriptors provide more than the evolution of our universe. They also tell us what will be the evolution when we start with other initial states.

In my view, the most significant disadvantage of this approach is that it loses  {\it gedanken measurability}. In the language of Raymond-Robichaud (2021), the local description is based on "noumenal" states. Given an ensemble of universes in a particular quantum Schr\"odinger state at a specific moment, and using external, omnipotent quantum devices, we could perform tomography to find the quantum state of the universe. In contrast, access to the ensemble of universes at a single point in time is insufficient to specify the system using descriptors; rather, we would need access throughout the universe's entire history. Furthermore, even with that historical access, we would still be unable to identify all the descriptors. This is because descriptors provide more than just the evolution of our actual universe—they also dictate what the evolution would be if starting from different initial states.

I encounter another difficulty with this approach. I cannot fully understand their definition of the concept of a world. Kuypers and Deutsch (2021) define worlds as ``autonomously evolving parts of reality, often called ‘parallel universes’$^1$'', adding a footnote: ``$^1$ Despite the name, an Everettian ‘parallel universe’ only ever has a finite spatial extent, because Everettian multiplicity arises from, and spreads through, local interactions.'' I do not see how a superposition of these finite `worlds' can represent the universe through the basic decomposition defined in Vaidman (2021):
\begin{align}\label{Universe}
  |\Psi\rangle_{\rm UNIVERSE}=\sum \alpha_i|\Psi\rangle_{{\rm WORLD}~i} ~.~~
\end{align}
 Although this decomposition is not defined precisely by physical parameters, since ``macroscopic'' and ``well localized'' are concepts specified by agents to help explaining their experience, it provides a description which is not less satisfactory than standard quantum approach, in which each world corresponds to some definite outcomes of all quantum measurements.

\section{Conclusions}

I reviewed various approaches to the question of information transfer in the process of quantum teleportation, a controversial topic that has not reached consensus. I argued that the basis of the explanation is the nonlocality of worlds in the MWI: Alice's local Bell measurement splits the world in different ways depending on the quantum state she receives to teleport. The operation creates worlds with information about Alice's qubit in Bob's (far away) location. This information, however, cannot be used to create the qubit without the information about identity of the world. 

The controversy about information transfer arises from the lack of a precise definition of the concept of information and the difficulties related to the quantum measurement problem. If we accept the reality of collapse in quantum measurements, then the teleportation procedure demonstrates a very problematic action at a distance. A local property in one location is changed immediately by action at remote location. The MWI, which does not contain action at a distance, does provide a coherent framework for discussing this problem; however, to achieve a clear picture, it is necessary to carefully consider the many-world structure of the physical universe. A useful concept of information has to be considered within a world, because information is used by agents, and every agent lives in a single world.  This concept of information should not be confused with a locally created pattern of the wavefunction of the universe which incorporates all parallel worlds together that propagates in space-time with a speed of light. 

The apparent contradiction with the information transfer picture of Deutsch and Hayden follows from different questions which were asked: a local description with descriptors describes not just the situation at a particular moment but also information about the past. And the locality of this description relies on a (strong) assumption about the past which cannot be verified at a particular time, even if we are given super-technology and an ensemble of universes. It seems that what is missing for a better comparison between the current and the Deutsch and Hayden approaches is a clear definition of a world in the Deutsch and Hayden picture, which I view as an important open question.

The only nonlocality of MWI is the nonseparability of entanglement. The paradox of transmitting an object which requires much more information than is transferred between Alice and Bob is explained by the presence of all qubits in Bob's site even before the teleportation. The  local Bell measurement spits the nonlocal world into the mixture of four worlds in every one of which Bob's particle has the information about the qubit that can be retrieved based on the identity of the world, the only information remained to be sent. 
This elegant picture of information transfer in the teleportation procedure provides, in my view, strong support for the MWI, especially relative to collapse theories: it avoids randomness (see Vaidman 2014) and action at distance (see Vaidman 2015). 
The local picture of Deutsch and Hayden (2000) avoids, in addition, the nonseparability, but for a high price in complexity of the new ontology.

This work has been supported in part by the Israel Science Foundation Grants No.~1713/24 and Alexander von Humboldt-Stiftung.

\vskip6pt

\enlargethispage{20pt}

\bibliographystyle{plainnat}

\end{document}